\documentclass[%
  reprint,
  showpacs,preprintnumbers,
  amsmath,amssymb,
  aps,
prl,
]{revtex4-1}

\usepackage{graphicx}
\usepackage{dcolumn}
\usepackage{bm}
\usepackage{hyperref}
\usepackage{multirow}
\usepackage{braket}
\usepackage{color}
\usepackage{ulem}

\usepackage{times}
\usepackage{booktabs}

\begin{document}

\preprint{APS/123-QED}

\title{
Spontaneous Antisymmetric Spin Splitting in Noncollinear Antiferromagnets \\without Spin-Orbit Coupling
}

\author{Satoru Hayami$^1$, Yuki Yanagi$^2$, and Hiroaki Kusunose$^3$}
\affiliation{
$^1$Department of Applied Physics, The University of Tokyo, Bunkyo, Tokyo 113-8656, Japan \\
$^2$Center for Computational Materials Science, Institute for Materials Research, Tohoku University, Sendai, Miyagi, 950-8577, Japan \\
$^3$Department of Physics, Meiji University, Kawasaki 214-8571, Japan 
}
 
\begin{abstract}
We propose a realization of an antisymmetric spin-split band structure through magnetic phase transitions without spin-orbit coupling. 
It enables us to utilize for a variety of magnetic-order-driven cross-correlated and nonreciprocal transport phenomena as similar to those in the spin-orbit-coupling oriented systems. 
We unveil its general condition as an emergence of a bond-type magnetic toroidal multipole (polar tensor) in the triangular unit with the noncollinear 120$^{\circ}$-AFM structures. 
By using the concept of augmented multipoles, we systematically analyze the phenomena in terms of an effective multipole coupling. 
Our multipole description is ubiquitously applied to any trigonal and hexagonal structures including the triangular, kagome, and breathing kagome structures, which provides how to design and engineer materials with a giant antisymmetric spin splitting and its physical responses even without the spin-orbit coupling. 
\end{abstract}
\maketitle

Antisymmetric spin splitting in electronic band structure, which is an opposite spin polarization at opposite wave vectors, has drawn considerable interest in noncentrosymmetric materials, since it is a fundamental origin of rich spintronic functionalities, nonreciprocal transports, and magneto-electric effects~\cite{Sinova_PhysRevLett.92.126603,furukawa2017observation,tokura2018nonreciprocal}.
It is typically realized in polar materials with the relatively large spin-orbit coupling (SOC), for instance, the nonmagnetic Rashba compound, BiTeI,~\cite{rashba1960properties,ishizaka2011giant,Bahramy_PhysRevB.84.041202} and monolayer transition-metal dichalcogenides, $MX_2$ ($M=$ Mo, W and $X=$ S, Se)~\cite{Zhu_PhysRevB.84.153402,wang2012electronics,ugeda2014giant,Andor_PhysRevX.4.011034}.

Even though a crystal structure is centrosymmetric, magnetic transition actualizes the antisymmetric spin splitting by an interplay between the kinetic motion of electrons and the magnetic structure via the SOC~\cite{tokura2014multiferroics,hayami2016emergent}. 
A spiral magnetic order is a typical example, which induces a linear magneto-electric effect in the presence of the nonzero vector spin chirality~\cite{Katsura_PhysRevLett.95.057205,Mostovoy_PhysRevLett.96.067601,Bulaevskii_PhysRevB.78.024402}. 
Another example is found in CoNb$_3$S$_6$ and Co$_4$Nb$_2$O$_9$ showing giant anomalous Hall and angle-dependent magneto-electric effects~\cite{Khanh_PhysRevB.93.075117,Khanh_PhysRevB.96.094434,Yanagi_PhysRevB.97.020404,ghimire2018large,li2019quantum,vsmejkal2019crystal}, respectively.
It is emphasized that the emergent antisymmetric spin splitting through the magnetic phase transition is more flexibly controllable, i.e., the spin splitting driven by magnetic orders can be varied or even switched on and off by external fields, pressure and temperature.
The complex interplay can be understood in a transparent manner by introducing the concept of the augmented multipole~\cite{suzuki2018first,Hayami_PhysRevB.98.165110,Watanabe_PhysRevB.98.245129}.

Since the above fascinating phenomena usually rely on the presence of the SOC, candidate materials are limited to those constituted by moderately heavier elements in a crystal structure under low space-group symmetry.
Such a limitation motivates a search for alternative mechanism to exhibit spin splitting without relying on the SOC.
This can be done by considering appropriate magnetic structures,
which break crystalline symmetry in addition to the time-reversal symmetry~\cite{Bulaevskii_PhysRevB.78.024402,zhang2018spin,hayami2019momentum}.
For example, a collinear-type antiferromagnetic (AFM) phase transition
in a nonsymmorphic organic compound, $\kappa$-(BETD-TTF)$_2$Cu[N(CN)$_2$]Cl~\cite{naka2019spin}, and a distorted tetragonal compound, RuO$_2$~\cite{Berlijn_PhysRevLett.118.077201,Ahn_PhysRevB.99.184432},  
result in the spin-current generation.
However, in the absence of the SOC it is proven that the collinear magnetic order leads only to the \textit{symmetric} spin splitting in momentum space even with the broken spatial inversion symmetry due to SU(2) symmetry in spin space~\cite{naka2019spin,hayami2019momentum}. 

In the present study, we propose a realization of \textit{antisymmetric} spin splitting by focusing on the triangular unit with the noncollinear 120$^{\circ}$-AFM structures, and clarify microscopic conditions for the emergent spin splitting from a general point of view by introducing the multipole description~\cite{hayami2018microscopic,Hayami_PhysRevB.98.165110}.
The condition we found is that the magnetic toroidal (MT) multipoles present in the hopping Hamiltonian and they couple with the noncollinear AFM order parameters within the same irreducible representation in the high-temperature series expansion.
We also predict possible cross-correlated and nonreciprocal transport phenomena in terms of an effective coupling among the multipole degrees of freedom, which can be modified by an external magnetic field for instance. 

Our multipole description is ubiquitously applied to any trigonal and hexagonal structures including the triangular, kagome, and breathing kagome structures. 
Our proposal is demonstrated for the trigonal noncollinear AFM Ba$_{3}$MnNb$_{2}$O$_{9}$ based on the density-functional-theory (DFT) calculation. 
The present mechanism provides potentially gigantic antisymmetric spin splitting due to its kinetic-motion origin without the SOC, which can be directly detected in spin- and angle-resolved photoemission spectroscopy.

\begin{figure}[t!]
\begin{center}
\includegraphics[width=1.0 \hsize]{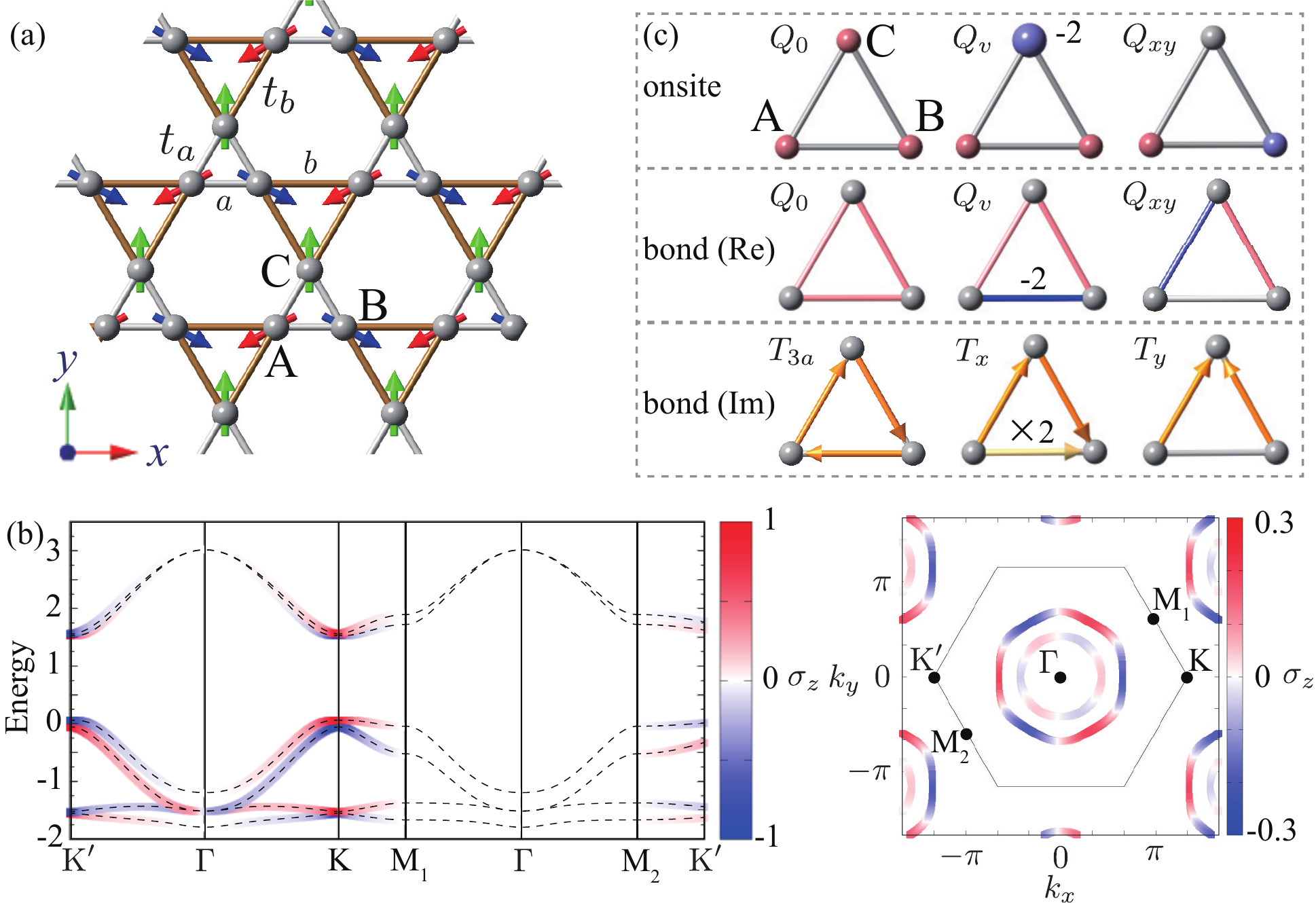} 
\caption{
\label{Fig:BKL_band}
(a) Noncollinear 120$^{\circ}$-AFM pattern in the breathing kagome structure. 
(b) (left panel) The band structure of the model in Eq.~(\ref{eq:Ham_BKL}) at $t_a = 1$, $t_b=0.5$, and $m=0.3$ and (right panel) the isoenergy surfaces at $\mu=-1$. 
The dashed lines show the band dispersions and the colormap shows the spin polarization of the $z$ components at each wave vector.
(c) Schematic pictures of the triangular-unit multipoles.
The red (blue) circles represent the positive (negative) onsite potential, and the red (blue) lines and orange arrows on each bond represent the positive (negative) real and imaginary hoppings, respectively. 
The gray lines represent no hoppings.
}
\end{center}
\end{figure}

We start by considering a breathing kagome system with the noncollinear 120$^{\circ}$-AFM structure in Fig.~\ref{Fig:BKL_band}(a), which is an intuitive example showing an antisymmetric spin splitting. 
The positions of the three sublattice sites are defined by $\bm{r}_{\rm A}= (0,0,0)$, $\bm{r}_{\rm B}= a(1,0,0)$, and $\bm{r}_{\rm C}= a(1/2,\sqrt{3}/2,0)$ and the lattice constant $a+b$ is set to be unity.
The tight-binding Hamiltonian is given by  
\begin{align}
\mathcal{H}=\Bigl( t_a \sum_{\sigma,\braket{ij}}^{\in \triangle} + t_b \sum_{\sigma,\braket{ij}}^{\in \bigtriangledown} \Bigr) c_{i\sigma}^{\dagger}c_{j\sigma}^{} + \sum_{i\sigma \sigma'} \bm{m}_i \cdot 
c_{i\sigma}^{\dagger} \bm{\sigma}_{\sigma \sigma'}c_{i\sigma'}^{},
\label{eq:Ham_BKL}
\end{align}
where $c^{\dagger}_{i\sigma}$ ($c_{i\sigma}^{}$) is the creation (annihilation) operator for site $i$ and spin $\sigma=\uparrow, \downarrow$. 
The first term represents the hoppings within upward triangles $t_a$ and downward triangles $t_b$. 
The second term represents the mean-field term corresponding to the magnetic order. 
We assume the noncollinear three-sublattice 120$^{\circ}$-AFM structure in the $xy$ plane with the order-parameter amplitude $m$, i.e., $\bm{m}_{\rm A}= m(-\sqrt{3}/2,-1/2,0)$, $\bm{m}_{\rm B}= m(\sqrt{3}/2,-1/2,0)$, and $\bm{m}_{\rm C}= m(0,1,0)$ in Fig.~\ref{Fig:BKL_band}(a), which can be naturally stabilized by the frustrated exchange interactions in the triangle unit, e.g., in the single-band Hubbard model on a triangular lattice~\cite{Yoshioka_PhysRevLett.103.036401}. 
We also consider the presence of the implicit small magnetic anisotropy due to magnetic dipole-dipole interactions and/or the SOC for the stabilization of the in-plane 120$^{\circ}$-AFM structure. 
Regardless of these stabilization mechanisms, the following properties are accounted for by the simple model in Eq.~(\ref{eq:Ham_BKL}). 

Figure~\ref{Fig:BKL_band}(b) shows the band structure at $t_a=1$, $t_b=0.5$, and $m=0.3$ where the colormap shows the spin polarization of the $z$ component. 
The results clearly exhibit the antisymmetric $z$-component spin polarization despite the AFM structure in the $xy$ plane; the spin polarization along the $\Gamma$-K  line is opposite to that along the $\Gamma$-K' line, while no spin polarization is found in the $\Gamma$-M$_{1,2}$ line.
The isoenergy surfaces at the chemical potential $\mu=-1$ in the right panel in Fig.~\ref{Fig:BKL_band}(b) indicate that the antisymmetric spin polarization keeps the three-fold rotational symmetry. 

This threefold out-of-plane antisymmetric spin splitting has close resemblance to that observed in the monolayer dicalcogenides with the SOC, which is so-called the Ising-type spin splitting~\cite{xiao2012coupled,mak2012control,zeng2012valley}.
However, the microscopic origin is totally different, i.e., the present case is the non-SOC origin, although the resultant antisymmetric spin splitting becomes a source of various cross-correlated and transport phenomena, such as the magneto-electric effect and nonreciprocal transport, as discussed below. 
The out-of-plane antisymmetric spin splitting can be detected by using spin- and angle-resolved photoemission spectroscopy~\cite{suzuki2014valley}.

The origin of the antisymmetric spin splitting can be intuitively captured by applying the multipole description to the model in Eq.~(\ref{eq:Ham_BKL})~\cite{Hayami_PhysRevB.98.165110,hayami2019momentum}, as the type of an additional crystalline symmetry breaking to the time-reversal symmetry is essential.
To demonstrate this, first we introduce the triangular unit with three sublattices A-C, and define the 9 multipole degrees of freedom as shown in Fig.~\ref{Fig:BKL_band}(c). 
Then, the spinless hopping matrix in the three-sublattice basis can be spanned by these multipoles.
Moreover, the mean-field magnetic structure is also described by the multipoles, which are known as the cluster multipoles~\cite{Suzuki_PhysRevB.95.094406,Suzuki_PhysRevB.99.174407}.

\begin{table*}[tb!]
\caption{
Multipole degrees of freedom in the triangular unit. 
The onsite potentials and nearest-neighbor hoppings are described by these multipoles. 
$\hat{\lambda}_\alpha$ ($\alpha=0$-$8$) are the Gell-Mann matrices.
We use the abbreviated notations, $\tilde{k}_{x}=k_{x}/2$, $\tilde{k}_{y}=\sqrt{3}k_{y}/2$, $p_{a}=t_{a}$, and $p_{b}=-t_{b}$.
}
\label{tab_multipoles1}
\centering
\begin{tabular}{ccccc} \hline\hline
electric & $\hat{Q}_{0}$ & $\hat{Q}_{v}$ & $\hat{Q}_{xy}$ \\ \hline
onsite & $\frac{1}{\sqrt{3}}\hat{\lambda}_0$ & $\frac{1}{\sqrt{2}}\hat{\lambda}_8$ & $\frac{1}{\sqrt{2}}\hat{\lambda}_3$ \\ 
real bond & $\frac{1}{\sqrt{6}}(\hat{\lambda}_1+\hat{\lambda}_4+\hat{\lambda}_6)$ & $\frac{\sqrt{3}}{6}(\hat{\lambda}_4+\hat{\lambda}_6-2\hat{\lambda}_1)$ & $\frac{1}{2}(-\hat{\lambda}_4+\hat{\lambda}_6)$ \\
form factor & $\sqrt{\frac{2}{3}}\sum_{\eta}t_{\eta}(\cos k_{x}\eta+2\cos\tilde{k}_{x}\eta\cos\tilde{k}_{y}\eta)$ & $\frac{2}{\sqrt{3}}\sum_{\eta}t_{\eta}(\cos\tilde{k}_{x}\eta\cos\tilde{k}_{y}\eta-\cos k_{x}\eta)$ & $2\sum_{\eta}t_{\eta}\sin\tilde{k}_{x}\eta\sin\tilde{k}_{y}\eta$ \\
\hline
magnetic & $\hat{T}_{3a}$ & $\hat{T}_{x}$ & $\hat{T}_{y}$  \\ \hline
imaginary bond & $\frac{1}{\sqrt{6}}(\hat{\lambda}_2-\hat{\lambda}_5+\hat{\lambda}_7)$ & $\frac{\sqrt{3}}{6}(\hat{\lambda}_7-\hat{\lambda}_5-2\hat{\lambda}_2)$ & $-\frac{1}{2}(\hat{\lambda}_5+\hat{\lambda}_7)$\\ 
form factor & $-\sqrt{\frac{2}{3}}\sum_{\eta}p_{\eta}(\sin k_x\eta-2\sin\tilde{k}_{x}\eta\cos\tilde{k}_{y}\eta)$ & $\frac{2}{\sqrt{3}}\sum_{\eta}p_{\eta}(\sin k_{x}\eta+\sin\tilde{k}_{x}\eta\cos\tilde{k}_{y}\eta)$ & $2\sum_{\eta}p_{\eta}\cos\tilde{k}_{x}\eta\sin\tilde{k}_{y}\eta$ \\
\hline
\hline
\end{tabular}
\end{table*}

The Hamiltonian in Eq.~(\ref{eq:Ham_BKL}) is Fourier transformed with respect to the unit cell as 
\begin{align}
\label{eq:Ham_mul}
\mathcal{H}=\!\! \sum_{\bm{k} \sigma \sigma' ll'}
c^{\dagger}_{\bm{k}l \sigma} \left[\delta_{\sigma \sigma'}(H_t^Q+H_t^T)^{ll'}+ \delta_{ll'} H_m^{\sigma\sigma'}\right]c_{\bm{k}l'\sigma'}, 
\end{align}
where $c^{\dagger}_{\bm{k}l \sigma}$ ($c_{\bm{k}l \sigma}$) is the Fourier transform of $c^{\dagger}_{i\sigma}$ ($c_{i\sigma}^{}$) at wave vector $\bm{k}$ and sublattice $l$. 
$H_t^Q$ and $H_t^T$ stand for the real and imaginary hopping matrices, respectively, which comes from the first term in Eq.~(\ref{eq:Ham_BKL}), and $H_m$ are the mean-field matrix from the second term in Eq.~(\ref{eq:Ham_BKL}). 
The matrices $H_t^Q$, $H_t^T$, and $H_m$ are decomposed in terms of the triangular-unit multipoles defined in Table~\ref{tab_multipoles1} as
\begin{align}
&H^{Q}_{t}=
Q_{0}(\bm{k})\hat{Q}_{0}^{(1)}+Q_{v}(\bm{k})\hat{Q}_{v}^{(1)}+Q_{xy}(\bm{k})\hat{Q}_{xy}^{(1)},
\cr&
H^{T}_{t}=T_{3a}(\bm{k})\hat{T}_{3a}^{(1)}+T_{x}(\bm{k})\hat{T}_{x}^{(1)}+T_{y}(\bm{k})\hat{T}_{y}^{(1)},
\cr&
H_{m}=-m (\hat{Q}_{xy}^{(0)}\hat{\sigma}_x +\hat{Q}_{v}^{(0)}\hat{\sigma}_y).
\label{eq:hammmul}
\end{align}

Here, three onsite potentials and three bonds with real hoppings are described by the linear combination of the electric monopole $\hat{Q}^{(n)}_0$ and two electric quadrupoles ($\hat{Q}^{(n)}_v, \hat{Q}^{(n)}_{xy}$), whereas three bonds with imaginary hoppings by two MT dipoles ($\hat{T}^{(1)}_x, \hat{T}^{(1)}_{y}$) and a MT octupole $\hat{T}^{(1)}_{3a}$, where the superscripts $n=0$ and $1$ stand for onsite and bond indices, respectively~\cite{Hayami_PhysRevB.98.165110,hayami2019momentum}. 
Note that the MT multipoles are defined as the bond degree of freedom in contrast to the conventional definition by the vector products of the spin (orbital) angular momentum and the position vector, where both definitions share the same symmetry properties.~\cite{EdererPhysRevB.76.214404,Spaldin_0953-8984-20-43-434203,Hayami_PhysRevB.90.024432,Gao_PhysRevB.97.134423}.
We use the standard Gell-Mann matrices to express each multipole in Table~\ref{tab_multipoles1}, and their schematic pictures are shown in Fig.~\ref{Fig:BKL_band}(c). 
Each multipole is normalized as ${\rm Tr} [\hat{X}^2] =1$.  
By using the molecular-orbital basis in the triangular unit~\cite{Hayami_PhysRevLett.122.147602,hayami2019momentum}, we identify the symmetry of each multipoles as indicated by the subscript.
The linear coefficients, the electric and MT multipoles $Q(\bm{k})$ and $T(\bm{k})$, represent the form factors, which are even and odd functions of $\bm{k}$, respectively.
Note that their $\bm{k}$ dependences are consistent with the general definition of multipoles in momentum representation~\cite{Hayami_PhysRevB.98.165110}.

In the multipole description, the active odd-rank MT bond multipoles (imaginary hoppings) can become the origin of the antisymmetric spin splitting, once the effective coupling between them and the mean-field multipoles is activated under spontaneous magnetic orders~\cite{comment_symmetric_SS}. 
Such an effective coupling is systematically obtained from the high-temperature expansion of the quantity at wave vector $\bm{k}$, 
\begin{align}
\label{eq:expec_spin}
\mathrm{Tr}[e^{-\beta \hat{\mathcal{H}}_{\bm{k}}} \hat{\sigma}_\mu]
= \sum_{s} \frac{(-\beta)^{s}}{s!}
g_s^{\mu}(\bm{k}), 
\end{align} 
where $\mu=0,x,y,z$, $\hat{\mathcal{H}}=\sum_{\bm{k}}\hat{\mathcal{H}}_{\bm{k}}$ and $\beta$ is the inverse temperature.
By means of the $s$-th order expansion coefficient of the $\mu$-component, $g_s^{\mu}(\bm{k})$, the corresponding effective multipole coupling is given by $g_{s}^{\mu}(\bm{k})\hat{\sigma}_{\mu}/2$.

The contribution to the antisymmetric spin splitting for the $z$-component is obtained at the 5-th order in Eq.~(\ref{eq:expec_spin}) as
\begin{align}
\label{eq:spinsplit_asym}
&g_{5}^{z}(\bm{k})
=\sqrt{\frac{2}{3}}m^2 \biggl\{Q_0(\bm{k}) [Q_{xy}(\bm{k}) T_y(\bm{k})-Q_v(\bm{k}) T_x(\bm{k})]
\cr&\quad
+ Q^2_0(\bm{k}) T_{3a}(\bm{k}) + \frac{1}{3\sqrt{2}}T_x(\bm{k})[T^2_x(\bm{k})-3 T^2_y(\bm{k})]\biggr\}.
\end{align}
Around $\bm{k}=0$, Eq.~(\ref{eq:spinsplit_asym}) reduces to $-m^2  t_a t_b (t_a-t_b)  k_x (k_x^2-3 k_y^2)(a+b)^3 /2$, which captures the qualitative behavior of the antisymmetric spin splitting in Fig.~\ref{Fig:BKL_band}(b).
It provides microscopic ingredients about the antisymmetric spin splitting.
The first is that the giant antisymmetric spin splitting could occur in the strong interaction regime, since the mean field $m$ in Eq.~(\ref{eq:hammmul}) is proportional to the repulsive interaction in the Hubbard model. 
The second is that the spin splitting is proportional to the square of the order parameter $m^2$, which implies that the two spin components, i.e., the noncollinear spin structure, is necessary to induce the spin splitting. 
Moreover, $m^2$ dependence indicates the AFM domain formation is irrelevant to this spin splitting, although the opposite chirality reverses the sign in Eq.~(\ref{eq:spinsplit_asym}). 
The third is that the spin splitting occurs for $t_a \neq 0$, $t_b \neq 0$, and $t_a \neq t_b$: the breathing structure ($a$-, $b$-bond inequivalency) is important.

Furthermore, 
the effective multipole coupling in Eq.~(\ref{eq:spinsplit_asym}) is a source of multiferroic responses, since each multipole is related to specific response tensors~\cite{Hayami_PhysRevB.98.165110}. 
For example, the effective coupling $Q^2_0(\bm{k})T_{3a}(\bm{k})\hat{\sigma}_{z}\sim k_x (k_x^2-3 k_y^2)\hat{\sigma}_{z}$ in Eq.~(\ref{eq:spinsplit_asym}), implies that a spontaneous threefold rotational nonreciprocity is induced by a magnetic field along the $z$ direction if one divides it as $k_{x}(k_{x}^{2}-3k_{y}^{2})\times \hat{\sigma}_{z}$. 
Similarly, the spin current along the $x$ direction with the $z$-spin component is expected by the $(x^2-y^2)$-type strain field by dividing the effective coupling $Q_{v}(\bm{k})T_{x}(\bm{k})\hat{\sigma}_{z}$ as $ k_{x}\hat{\sigma}_{z}\times(k_{x}^{2}-k_{y}^{2})$.

The analyses are straightforwardly extended to include an external magnetic field with the Zeeman coupling, $- \bm{H}\cdot \sum_{i\sigma\sigma'}c_{i\sigma}^{\dagger} \bm{\sigma}_{\sigma \sigma'}c_{i\sigma'}^{}$, yielding a rich variety of band deformations depending on the field direction.
For $\bm{H}\parallel [100]$, the directional antisymmetric spin splitting with $k^5_y \hat{\sigma}_z$ is induced by the coupling between $\hat{T}_{3a}$, $\hat{Q}_{xy}$, and $\hat{\sigma}_z$. 
This band deformation describes the emergent magneto-electric (ME) effect where the electric polarization along the $x$ direction, $Q_x$, is induced by $H_x$, since $k^5_y \hat{\sigma}_z$ is the same symmetry as $Q_x$~\cite{Hayami_PhysRevB.98.165110}. 
In a similar way, different ME couplings are obtained for $\bm{H}\parallel [010]$ and $\bm{H}\parallel [001]$: the $k^5_x \hat{\sigma}_z$-type band deformation corresponding to $Q_y$ for $\bm{H}\parallel [010]$ and the $k^5_x \hat{\sigma}_y-k^5_y \hat{\sigma}_x$-type band deformation corresponding to $Q_z$ for $\bm{H}\parallel [001]$. 
Thus, the 120$^{\circ}$-AFM order in the breathing kagome system exhibits the longitudinal ME effect ($\bm{Q}\parallel \bm{H}$). 
Note that the qualitatively similar results are also obtained by the symmetry analysis based on the cluster multipole theory~\cite{Suzuki_PhysRevB.99.174407}, although our approach is apparent for microscopic conditions to induce the antisymmetric spin splitting in a systematic way.
We summarize the effective coupling and relevant responses under the magnetic fields in Table~\ref{tab_ziba}~\cite{comment_mpform}.

\begin{table}[t!]
\caption{
Effective multipole couplings under an external magnetic field~\cite{comment_mpform}. 
The lowest-order band deformations $\bm{g}_{s}(\bm{k})\cdot\hat{\bm{\sigma}}$, the wave-vector $\bm{k}$ dependences around $\bm{k}=0$, and relevant physical responses are shown, where $\bm{Q}$ and $\bm{T}$ are the electric polarization and magnetic toroidalization. 
ME and NR indicate magneto-electric and nonreciprocal responses, respectively.
}
\label{tab_ziba}
\centering
\begin{tabular}{ccccc} \hline\hline
$\bm{H}$ & $\bm{g}_{s}(\bm{k})\cdot \hat{\bm{\sigma}}$ & $\bm{k}\to0$ limit & response \\
 \hline
$[100]$   &  $m^3 H_x Q_{xy}(\bm{k})T_{3a}(\bm{k})\hat{\sigma}_z$ & $ k^5_y \hat{\sigma}_z\sim Q_{x}$  & ME \\
\hline
$[010]$   &   $m^3 H_y Q_{v}(\bm{k})T_{3a}(\bm{k})\hat{\sigma}_z$ & $ k^5_x \hat{\sigma}_z\sim Q_{y}$  & ME \\
\hline
\multirow{3}{*}{$[001]$} &   $m^2 H_z T_{3a}(\bm{k}) \hat{\sigma}_0(\bm{k})$  & $k_x (k_x^2-3k_y^2)\hat{\sigma}_{0}\sim T_{3a}$ & NR \\
 &   $m^3 H_z  (Q_{xy}(\bm{k}) T_{3a}(\bm{k}) \hat{\sigma}_x$ & \multirow{2}{*}{$k^5_x \hat{\sigma}_y
-k^5_y \hat{\sigma}_x\sim Q_{z} $}  & \multirow{2}{*}{ ME } \\
 &   $+Q_{v}(\bm{k}) T_{3a}(\bm{k})\hat{\sigma}_y)$ &  & \\
 \hline
$[011]$ & $m^3 H_y H_z  Q_{v}(\bm{k}) T_{3a}(\bm{k}) \hat{\sigma}_{0}$ & $ k^5_x \hat{\sigma}_{0}\sim T_{x}$ & NR \\
\hline
$[101]$ &   $m^3 H_x H_z  Q_{xy}(\bm{k}) T_{3a}(\bm{k})\hat{\sigma}_{0}$ & $ k^5_y \hat{\sigma}_{0}\sim T_{y}$  & NR \\
\hline\hline
\end{tabular}
\end{table}

\begin{figure}[t!]
\begin{center}
\includegraphics[width=1.0 \hsize]{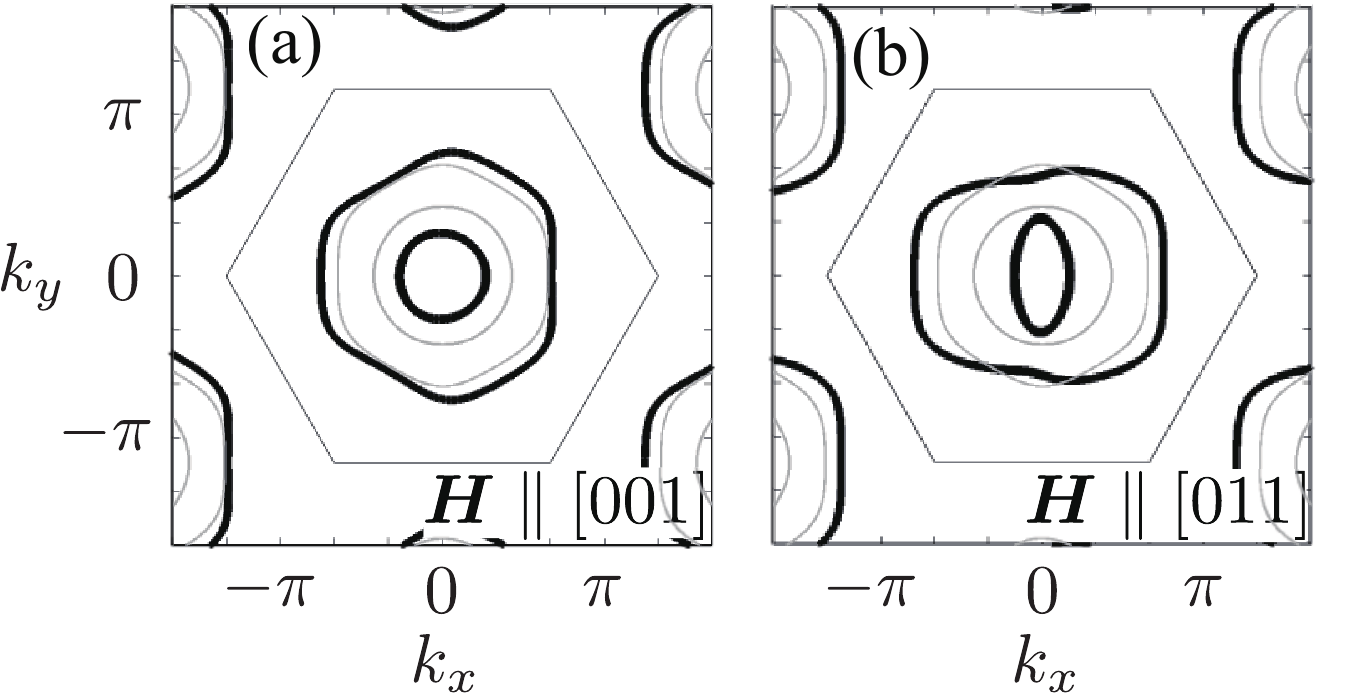} 
\caption{
\label{Fig:BKL_ziba}
The band deformations in the breathing kagome AFM at $|\bm{H}|=0.3$ along (a) [001] and (b) [011] directions.
The thin gray lines represent the isoenergy surfaces at $|\bm{H}|=0$. 
}
\end{center}
\end{figure}

Interestingly, spin-independent antisymmetric band deformations are realized when the magnetic field is applied along the $z$ direction as shown in Fig.~\ref{Fig:BKL_ziba}(a), where the effective multipole coupling is expressed as $m^2 H_z T_{3a}(\bm{k})\hat{\sigma}_{0} \sim m^2 H_z k_x(k_x^2-3k_y^2) \hat{\sigma}_{0}$. 
This type of antisymmetric band deformation becomes a microscopic source of the angle-dependent nonreciprocal transport.
Moreover, when $\bm{H}$ is rotated from [001] to [011], the additional contribution, $k_x^5 \hat{\sigma}_{0}$, appears due to the effective multipole coupling as $m^3 H_y H_z  Q_{v}(\bm{k}) T_{3a}(\bm{k})\hat{\sigma}_{0}$ [Fig.~\ref{Fig:BKL_ziba}(b)], which means that the magnetic field can induce the MT dipole, $T_{x}$.
Similar nonreciprocal dispersions have been studied in the localized spin model~\cite{Cheon_PhysRevB.98.184405,Maksimov_PhysRevX.9.021017}.

\begin{figure}[t!]
\begin{center}
\includegraphics[width=1.0 \hsize]{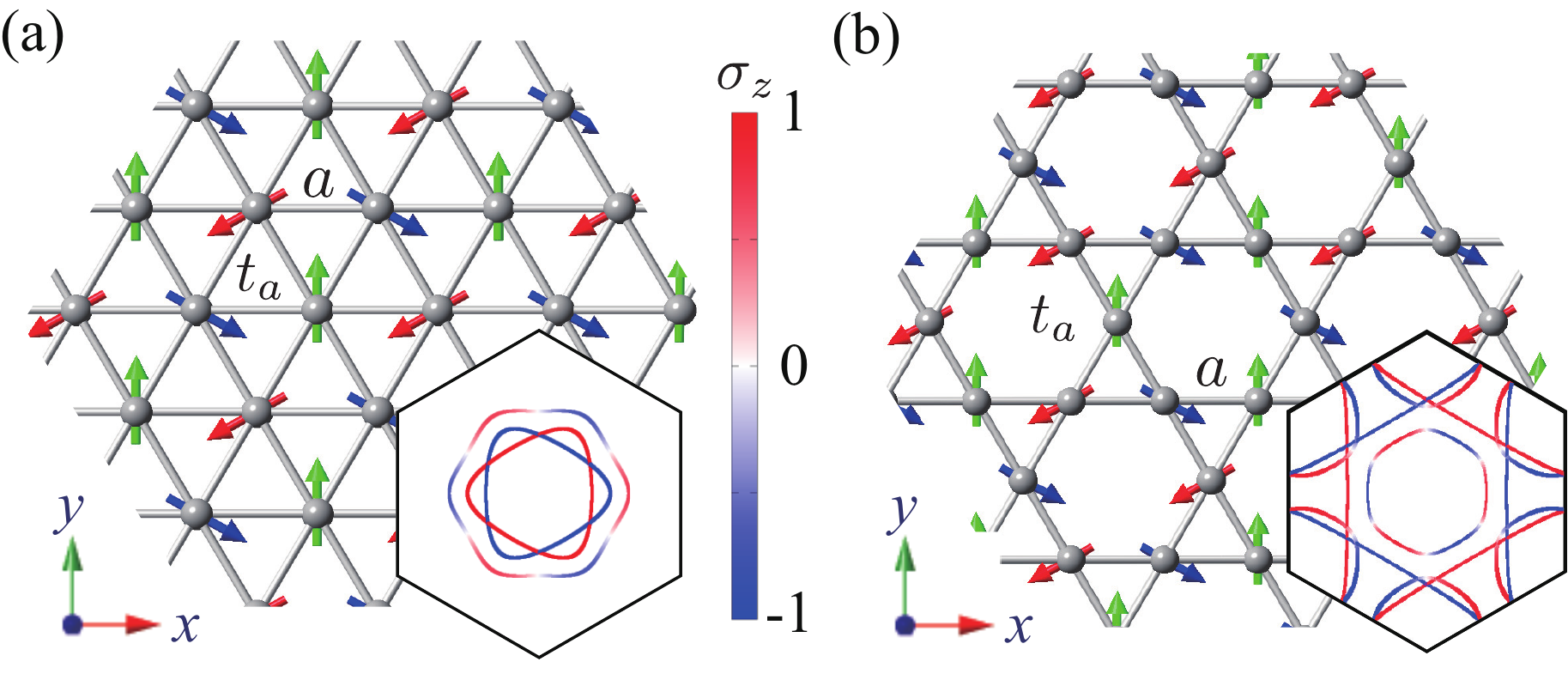} 
\caption{
\label{Fig:other}
Schematic pictures of the 120$^{\circ}$ AFM on (a) triangular and (b) kagome lattices with the $\sqrt{3}\times \sqrt{3}$ structures. 
In the inset, corresponding isoenergy surfaces where the contour shows the $z$-spin component are presented. 
The model parameters are given by (a) $t_a=1$, $m=0.5$, and $\mu=-2.5$ and (b) $t_a=1$, $m=0.5$, and $\mu=0$.
}
\end{center}
\end{figure}

\begin{figure}[t]
\begin{center}
\includegraphics[width=1.0 \hsize]{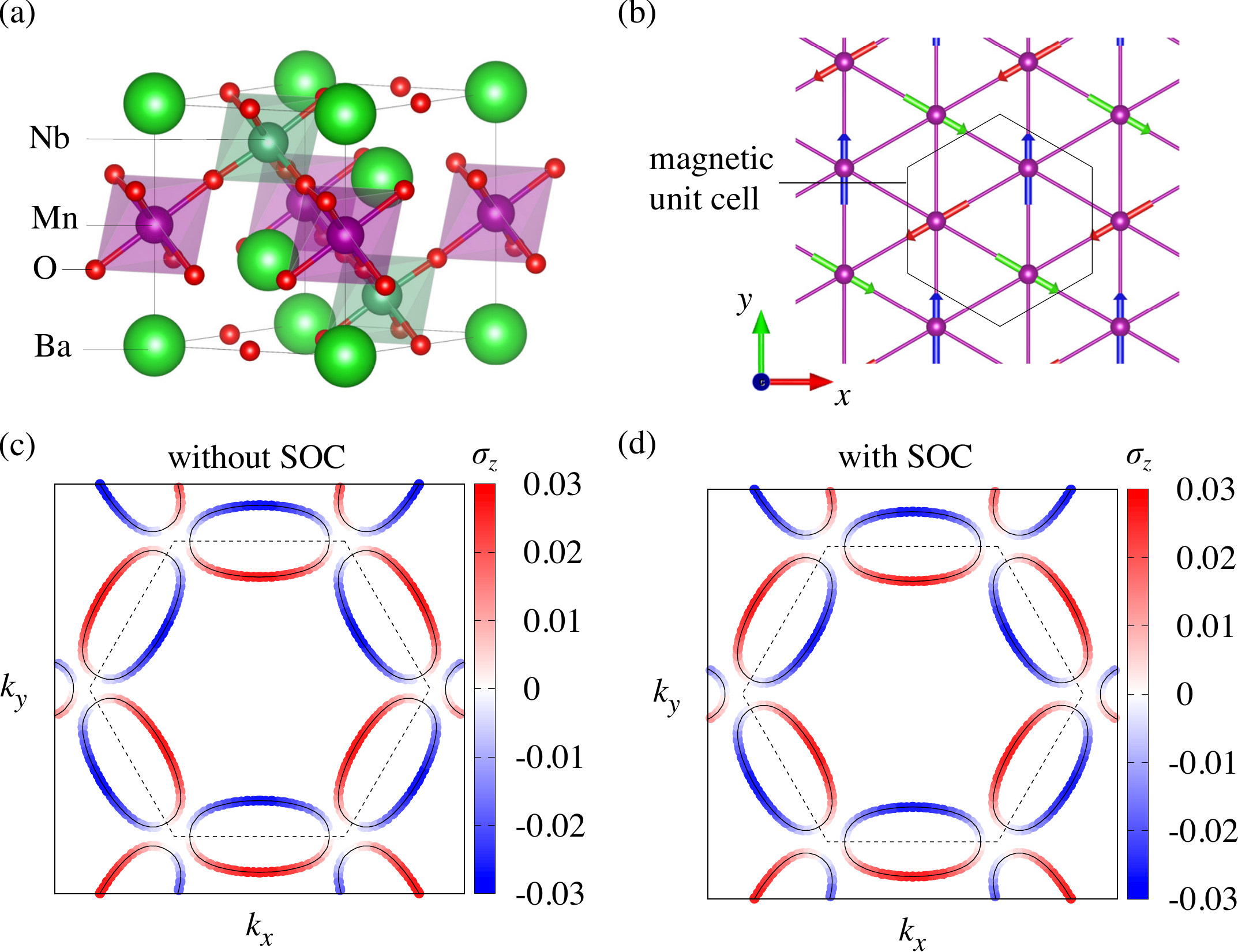}
\caption{(a) Crystal and (b) magnetic structures of Ba$_3$MnNb$_2$O$_9$.
The isoenergy surfaces on $k_z=0$ plane at $\mu= -0.05$ eV (c) without and (d) with the SOC in the AFM state, where the top of the valence band is set to $0$ eV. 
\label{Fig:Ba3MnNb2O9}
}
\end{center}
\end{figure}

So far, we have considered the specific breathing kagome structure. 
Similar analyses can be directly applied to any other systems with the triangular unit, such as the triangular and kagome systems.
For example, as the multipoles $\hat{Q}^{(1)}_0$ and $\hat{T}^{(1)}_{3a}$ are active in the three-sublattice 120$^\circ$-AFM order on a triangular lattice, the antisymmetric spin splitting is expected [Fig.~\ref{Fig:other}(a)]. 
Note that there are no additional antisymmetric band deformations induced by the magnetic field, since there are no active $\hat{Q}^{(1)}_{v}$ and $\hat{Q}^{(1)}_{xy}$ [see also Table~\ref{tab_ziba}]. 
The nearly 120$^\circ$-AFM materials, such as CsFeCl$_3$~\cite{Hayashida_PhysRevB.97.140405} and PdCrO$_2$~\cite{mekata1995magnetic,ghannadzadeh2017simultaneous}, are candidates to exhibit the antisymmetric spin splitting. 
In a similar way, the antisymmetric spin splitting is expected for the $\sqrt{3}\times \sqrt{3}$ AFM order on the simple kagome structure, where only $\hat{Q}^{(1)}_0$ and $\hat{T}^{(1)}_{3a}$ are active multipoles as shown in Fig.~\ref{Fig:other}(b).
Moreover, the lower-symmetry trigonal material also shows spin-split physics, such as trimer and triangular tube magnets, LuFeO$_3$~\cite{Foggetti_PhysRevB.100.180408} and CsCrF$_4$~\cite{Seki_PhysRevB.91.224403,hagihala2019magnetic}, which possesses the same active multipoles as the breathing kagome structure.

Finally, we demonstrate the emergent antisymmetric spin splitting in Ba$_{3}$MnNb$_{2}$O$_{9}$~\cite{Lee_PhysRevB.90.224402}.
This compound belongs to the trigonal space group $P\bar{3}m1$ (No. 164), and the high-spin state ($S=5/2$) of Mn$^{2+}$ ions exhibits the $120^{\circ}$-AFM structure with out-of-plane cantings on the triangular lattice at low temperatures as shown in Figs.~\ref{Fig:Ba3MnNb2O9}(a) and (b)~\cite{Lee_PhysRevB.90.224402}.
We calculate the expected 
 AFM band structures of  Ba$_{3}$MnNb$_{2}$O$_{9}$ with and without the SOC based on the DFT with the  generalized gradient approximation plus  $U$ method~\cite{Perdew_PhysRevLett.77.3865,Dudarev_PhysRevB.57.1505} by using the Vienna \textit{ab initio} simulation package (\textsc{vasp})~\cite{Kresse_PhysRevB.54.11169,vasp}, where we employ the projector augmented wave (PAW) potentials~\cite{Blochl_PhysRevB.50.17953,Kresse_PhysRevB.59.1758} and set $U=3.0$ eV for Mn-$3d$-orbitals according to the previous study~\cite{Lee_PhysRevB.90.224402}.  
Figure~\ref{Fig:Ba3MnNb2O9}(c) shows the isoenergy surfaces without the SOC projected onto the $\sigma_z$-component in the AFM state at zero magnetic field.
The results are consistent with the analysis in the simple triangular AFM in Fig.~\ref{Fig:other}(a), i.e., the antisymmetric $z$-spin polarization in the form of $k_y(3k^2_x-k_y^2)$. 
Note that the SOC for the Mn-$3d$-orbitals is small and does not have any significant impact on the antisymmetric spin-splitting as shown in Fig.~\ref{Fig:Ba3MnNb2O9}(d).  
In addition, we also confirmed that the isoenergy surfaces are deformed antisymmetrically for $\bm{H}\parallel [001]$.
Therefore, this compound can be an archetypal example of the antisymmetric SOC physics induced by a noncollinear magnetic ordering.

In summary, we clarify general conditions for emergent antisymmetric spin-split band structures in noncollinear magnets. 
The following three conditions are enough to obtain the {\it antisymmetric} spin splitting in the band structure without the SOC: 
(1) the triangular unit with the 120$^{\circ}$-AFM structure, (2) inversion symmetry breaking and (3) active MT multipoles (imaginary hopping) in the one-body Hamiltonian. 
We also demonstrate the origin of the cross-correlated coupling and nonreciprocal transport is attributed to the effective microscopic multipole couplings.
As our analysis on the basis of the multipole description is ubiquitously applied to any systems with the triangular unit, the result will shed light on potential candidate materials with a giant spin splitting even without the SOC.

\begin{acknowledgments}
This research was supported by JSPS KAKENHI Grants Numbers JP15H05885, JP18H04296 (J-Physics), JP18K13488, JP19K03752, JP19H01834, and JP20K05299. 
This work was also supported by the Toyota Riken Scholarship. 
DFT calculations were performed using the MAterial science Supercomputing system for Advanced MUlti-scale simulations towards NExt-generation-Institute for Materials Research (MASAMUNE- IMR) of the Center for Computational Materials Science, Institute for Materials Research, Tohoku University. 
\end{acknowledgments}

\bibliographystyle{apsrev}
\bibliography{ref}

\begin{thebibliography}{59}
\expandafter\ifx\csname natexlab\endcsname\relax\def\natexlab#1{#1}\fi
\expandafter\ifx\csname bibnamefont\endcsname\relax
  \def\bibnamefont#1{#1}\fi
\expandafter\ifx\csname bibfnamefont\endcsname\relax
  \def\bibfnamefont#1{#1}\fi
\expandafter\ifx\csname citenamefont\endcsname\relax
  \def\citenamefont#1{#1}\fi
\expandafter\ifx\csname url\endcsname\relax
  \def\url#1{\texttt{#1}}\fi
\expandafter\ifx\csname urlprefix\endcsname\relax\def\urlprefix{URL }\fi
\providecommand{\bibinfo}[2]{#2}
\providecommand{\eprint}[2][]{\url{#2}}

\bibitem[{\citenamefont{Sinova et~al.}(2004)\citenamefont{Sinova, Culcer, Niu,
  Sinitsyn, Jungwirth, and MacDonald}}]{Sinova_PhysRevLett.92.126603}
\bibinfo{author}{\bibfnamefont{J.}~\bibnamefont{Sinova}},
  \bibinfo{author}{\bibfnamefont{D.}~\bibnamefont{Culcer}},
  \bibinfo{author}{\bibfnamefont{Q.}~\bibnamefont{Niu}},
  \bibinfo{author}{\bibfnamefont{N.~A.} \bibnamefont{Sinitsyn}},
  \bibinfo{author}{\bibfnamefont{T.}~\bibnamefont{Jungwirth}},
  \bibnamefont{and} \bibinfo{author}{\bibfnamefont{A.~H.}
  \bibnamefont{MacDonald}}, \bibinfo{journal}{Phys. Rev. Lett.}
  \textbf{\bibinfo{volume}{92}}, \bibinfo{pages}{126603}
  (\bibinfo{year}{2004}).

\bibitem[{\citenamefont{Furukawa et~al.}(2017)\citenamefont{Furukawa,
  Shimokawa, Kobayashi, and Itou}}]{furukawa2017observation}
\bibinfo{author}{\bibfnamefont{T.}~\bibnamefont{Furukawa}},
  \bibinfo{author}{\bibfnamefont{Y.}~\bibnamefont{Shimokawa}},
  \bibinfo{author}{\bibfnamefont{K.}~\bibnamefont{Kobayashi}},
  \bibnamefont{and} \bibinfo{author}{\bibfnamefont{T.}~\bibnamefont{Itou}},
  \bibinfo{journal}{Nat. Commun.} \textbf{\bibinfo{volume}{8}},
  \bibinfo{pages}{954} (\bibinfo{year}{2017}).

\bibitem[{\citenamefont{Tokura and Nagaosa}(2018)}]{tokura2018nonreciprocal}
\bibinfo{author}{\bibfnamefont{Y.}~\bibnamefont{Tokura}} \bibnamefont{and}
  \bibinfo{author}{\bibfnamefont{N.}~\bibnamefont{Nagaosa}},
  \bibinfo{journal}{Nat. Commun.} \textbf{\bibinfo{volume}{9}},
  \bibinfo{pages}{1} (\bibinfo{year}{2018}).

\bibitem[{\citenamefont{Rashba}(1960)}]{rashba1960properties}
\bibinfo{author}{\bibfnamefont{E.~I.} \bibnamefont{Rashba}},
  \bibinfo{journal}{Sov. Phys. Solid State} \textbf{\bibinfo{volume}{2}},
  \bibinfo{pages}{1109} (\bibinfo{year}{1960}).

\bibitem[{\citenamefont{Ishizaka et~al.}(2011)\citenamefont{Ishizaka, Bahramy,
  Murakawa, Sakano, Shimojima, Sonobe, Koizumi, Shin, Miyahara, Kimura
  et~al.}}]{ishizaka2011giant}
\bibinfo{author}{\bibfnamefont{K.}~\bibnamefont{Ishizaka}},
  \bibinfo{author}{\bibfnamefont{M.}~\bibnamefont{Bahramy}},
  \bibinfo{author}{\bibfnamefont{H.}~\bibnamefont{Murakawa}},
  \bibinfo{author}{\bibfnamefont{M.}~\bibnamefont{Sakano}},
  \bibinfo{author}{\bibfnamefont{T.}~\bibnamefont{Shimojima}},
  \bibinfo{author}{\bibfnamefont{T.}~\bibnamefont{Sonobe}},
  \bibinfo{author}{\bibfnamefont{K.}~\bibnamefont{Koizumi}},
  \bibinfo{author}{\bibfnamefont{S.}~\bibnamefont{Shin}},
  \bibinfo{author}{\bibfnamefont{H.}~\bibnamefont{Miyahara}},
  \bibinfo{author}{\bibfnamefont{A.}~\bibnamefont{Kimura}},
  \bibnamefont{et~al.}, \bibinfo{journal}{Nat. mater.}
  \textbf{\bibinfo{volume}{10}}, \bibinfo{pages}{521} (\bibinfo{year}{2011}).

\bibitem[{\citenamefont{Bahramy et~al.}(2011)\citenamefont{Bahramy, Arita, and
  Nagaosa}}]{Bahramy_PhysRevB.84.041202}
\bibinfo{author}{\bibfnamefont{M.~S.} \bibnamefont{Bahramy}},
  \bibinfo{author}{\bibfnamefont{R.}~\bibnamefont{Arita}}, \bibnamefont{and}
  \bibinfo{author}{\bibfnamefont{N.}~\bibnamefont{Nagaosa}},
  \bibinfo{journal}{Phys. Rev. B} \textbf{\bibinfo{volume}{84}},
  \bibinfo{pages}{041202} (\bibinfo{year}{2011}).

\bibitem[{\citenamefont{Zhu et~al.}(2011)\citenamefont{Zhu, Cheng, and
  Schwingenschl\"ogl}}]{Zhu_PhysRevB.84.153402}
\bibinfo{author}{\bibfnamefont{Z.~Y.} \bibnamefont{Zhu}},
  \bibinfo{author}{\bibfnamefont{Y.~C.} \bibnamefont{Cheng}}, \bibnamefont{and}
  \bibinfo{author}{\bibfnamefont{U.}~\bibnamefont{Schwingenschl\"ogl}},
  \bibinfo{journal}{Phys. Rev. B} \textbf{\bibinfo{volume}{84}},
  \bibinfo{pages}{153402} (\bibinfo{year}{2011}).

\bibitem[{\citenamefont{Wang et~al.}(2012)\citenamefont{Wang, Kalantar-Zadeh,
  Kis, Coleman, and Strano}}]{wang2012electronics}
\bibinfo{author}{\bibfnamefont{Q.~H.} \bibnamefont{Wang}},
  \bibinfo{author}{\bibfnamefont{K.}~\bibnamefont{Kalantar-Zadeh}},
  \bibinfo{author}{\bibfnamefont{A.}~\bibnamefont{Kis}},
  \bibinfo{author}{\bibfnamefont{J.~N.} \bibnamefont{Coleman}},
  \bibnamefont{and} \bibinfo{author}{\bibfnamefont{M.~S.}
  \bibnamefont{Strano}}, \bibinfo{journal}{Nat. Nanotech.}
  \textbf{\bibinfo{volume}{7}}, \bibinfo{pages}{699} (\bibinfo{year}{2012}).

\bibitem[{\citenamefont{Ugeda et~al.}(2014)\citenamefont{Ugeda, Bradley, Shi,
  Felipe, Zhang, Qiu, Ruan, Mo, Hussain, Shen et~al.}}]{ugeda2014giant}
\bibinfo{author}{\bibfnamefont{M.~M.} \bibnamefont{Ugeda}},
  \bibinfo{author}{\bibfnamefont{A.~J.} \bibnamefont{Bradley}},
  \bibinfo{author}{\bibfnamefont{S.-F.} \bibnamefont{Shi}},
  \bibinfo{author}{\bibfnamefont{H.}~\bibnamefont{Felipe}},
  \bibinfo{author}{\bibfnamefont{Y.}~\bibnamefont{Zhang}},
  \bibinfo{author}{\bibfnamefont{D.~Y.} \bibnamefont{Qiu}},
  \bibinfo{author}{\bibfnamefont{W.}~\bibnamefont{Ruan}},
  \bibinfo{author}{\bibfnamefont{S.-K.} \bibnamefont{Mo}},
  \bibinfo{author}{\bibfnamefont{Z.}~\bibnamefont{Hussain}},
  \bibinfo{author}{\bibfnamefont{Z.-X.} \bibnamefont{Shen}},
  \bibnamefont{et~al.}, \bibinfo{journal}{Nat. Mater.}
  \textbf{\bibinfo{volume}{13}}, \bibinfo{pages}{1091} (\bibinfo{year}{2014}).

\bibitem[{\citenamefont{Korm\'anyos et~al.}(2014)\citenamefont{Korm\'anyos,
  Z\'olyomi, Drummond, and Burkard}}]{Andor_PhysRevX.4.011034}
\bibinfo{author}{\bibfnamefont{A.}~\bibnamefont{Korm\'anyos}},
  \bibinfo{author}{\bibfnamefont{V.}~\bibnamefont{Z\'olyomi}},
  \bibinfo{author}{\bibfnamefont{N.~D.} \bibnamefont{Drummond}},
  \bibnamefont{and} \bibinfo{author}{\bibfnamefont{G.}~\bibnamefont{Burkard}},
  \bibinfo{journal}{Phys. Rev. X} \textbf{\bibinfo{volume}{4}},
  \bibinfo{pages}{011034} (\bibinfo{year}{2014}).

\bibitem[{\citenamefont{Tokura et~al.}(2014)\citenamefont{Tokura, Seki, and
  Nagaosa}}]{tokura2014multiferroics}
\bibinfo{author}{\bibfnamefont{Y.}~\bibnamefont{Tokura}},
  \bibinfo{author}{\bibfnamefont{S.}~\bibnamefont{Seki}}, \bibnamefont{and}
  \bibinfo{author}{\bibfnamefont{N.}~\bibnamefont{Nagaosa}},
  \bibinfo{journal}{Rep. Prog. Phys.} \textbf{\bibinfo{volume}{77}},
  \bibinfo{pages}{076501} (\bibinfo{year}{2014}).

\bibitem[{\citenamefont{Hayami et~al.}(2016)\citenamefont{Hayami, Kusunose, and
  Motome}}]{hayami2016emergent}
\bibinfo{author}{\bibfnamefont{S.}~\bibnamefont{Hayami}},
  \bibinfo{author}{\bibfnamefont{H.}~\bibnamefont{Kusunose}}, \bibnamefont{and}
  \bibinfo{author}{\bibfnamefont{Y.}~\bibnamefont{Motome}},
  \bibinfo{journal}{J. Phys.: Condens. Matter} \textbf{\bibinfo{volume}{28}},
  \bibinfo{pages}{395601} (\bibinfo{year}{2016}).

\bibitem[{\citenamefont{Katsura et~al.}(2005)\citenamefont{Katsura, Nagaosa,
  and Balatsky}}]{Katsura_PhysRevLett.95.057205}
\bibinfo{author}{\bibfnamefont{H.}~\bibnamefont{Katsura}},
  \bibinfo{author}{\bibfnamefont{N.}~\bibnamefont{Nagaosa}}, \bibnamefont{and}
  \bibinfo{author}{\bibfnamefont{A.~V.} \bibnamefont{Balatsky}},
  \bibinfo{journal}{Phys. Rev. Lett.} \textbf{\bibinfo{volume}{95}},
  \bibinfo{pages}{057205} (\bibinfo{year}{2005}).

\bibitem[{\citenamefont{Mostovoy}(2006)}]{Mostovoy_PhysRevLett.96.067601}
\bibinfo{author}{\bibfnamefont{M.}~\bibnamefont{Mostovoy}},
  \bibinfo{journal}{Phys. Rev. Lett.} \textbf{\bibinfo{volume}{96}},
  \bibinfo{pages}{067601} (\bibinfo{year}{2006}).

\bibitem[{\citenamefont{Bulaevskii et~al.}(2008)\citenamefont{Bulaevskii,
  Batista, Mostovoy, and Khomskii}}]{Bulaevskii_PhysRevB.78.024402}
\bibinfo{author}{\bibfnamefont{L.~N.} \bibnamefont{Bulaevskii}},
  \bibinfo{author}{\bibfnamefont{C.~D.} \bibnamefont{Batista}},
  \bibinfo{author}{\bibfnamefont{M.~V.} \bibnamefont{Mostovoy}},
  \bibnamefont{and} \bibinfo{author}{\bibfnamefont{D.~I.}
  \bibnamefont{Khomskii}}, \bibinfo{journal}{Phys. Rev. B}
  \textbf{\bibinfo{volume}{78}}, \bibinfo{pages}{024402}
  (\bibinfo{year}{2008}).

\bibitem[{\citenamefont{Khanh et~al.}(2016)\citenamefont{Khanh, Abe, Sagayama,
  Nakao, Hanashima, Kiyanagi, Tokunaga, and Arima}}]{Khanh_PhysRevB.93.075117}
\bibinfo{author}{\bibfnamefont{N.~D.} \bibnamefont{Khanh}},
  \bibinfo{author}{\bibfnamefont{N.}~\bibnamefont{Abe}},
  \bibinfo{author}{\bibfnamefont{H.}~\bibnamefont{Sagayama}},
  \bibinfo{author}{\bibfnamefont{A.}~\bibnamefont{Nakao}},
  \bibinfo{author}{\bibfnamefont{T.}~\bibnamefont{Hanashima}},
  \bibinfo{author}{\bibfnamefont{R.}~\bibnamefont{Kiyanagi}},
  \bibinfo{author}{\bibfnamefont{Y.}~\bibnamefont{Tokunaga}}, \bibnamefont{and}
  \bibinfo{author}{\bibfnamefont{T.}~\bibnamefont{Arima}},
  \bibinfo{journal}{Phys. Rev. B} \textbf{\bibinfo{volume}{93}},
  \bibinfo{pages}{075117} (\bibinfo{year}{2016}).

\bibitem[{\citenamefont{Khanh et~al.}(2017)\citenamefont{Khanh, Abe, Kimura,
  Tokunaga, and Arima}}]{Khanh_PhysRevB.96.094434}
\bibinfo{author}{\bibfnamefont{N.~D.} \bibnamefont{Khanh}},
  \bibinfo{author}{\bibfnamefont{N.}~\bibnamefont{Abe}},
  \bibinfo{author}{\bibfnamefont{S.}~\bibnamefont{Kimura}},
  \bibinfo{author}{\bibfnamefont{Y.}~\bibnamefont{Tokunaga}}, \bibnamefont{and}
  \bibinfo{author}{\bibfnamefont{T.}~\bibnamefont{Arima}},
  \bibinfo{journal}{Phys. Rev. B} \textbf{\bibinfo{volume}{96}},
  \bibinfo{pages}{094434} (\bibinfo{year}{2017}).

\bibitem[{\citenamefont{Yanagi et~al.}(2018)\citenamefont{Yanagi, Hayami, and
  Kusunose}}]{Yanagi_PhysRevB.97.020404}
\bibinfo{author}{\bibfnamefont{Y.}~\bibnamefont{Yanagi}},
  \bibinfo{author}{\bibfnamefont{S.}~\bibnamefont{Hayami}}, \bibnamefont{and}
  \bibinfo{author}{\bibfnamefont{H.}~\bibnamefont{Kusunose}},
  \bibinfo{journal}{Phys. Rev. B} \textbf{\bibinfo{volume}{97}},
  \bibinfo{pages}{020404} (\bibinfo{year}{2018}).

\bibitem[{\citenamefont{Ghimire et~al.}(2018)\citenamefont{Ghimire, Botana,
  Jiang, Zhang, Chen, and Mitchell}}]{ghimire2018large}
\bibinfo{author}{\bibfnamefont{N.~J.} \bibnamefont{Ghimire}},
  \bibinfo{author}{\bibfnamefont{A.}~\bibnamefont{Botana}},
  \bibinfo{author}{\bibfnamefont{J.}~\bibnamefont{Jiang}},
  \bibinfo{author}{\bibfnamefont{J.}~\bibnamefont{Zhang}},
  \bibinfo{author}{\bibfnamefont{Y.-S.} \bibnamefont{Chen}}, \bibnamefont{and}
  \bibinfo{author}{\bibfnamefont{J.}~\bibnamefont{Mitchell}},
  \bibinfo{journal}{Nat. Commun.} \textbf{\bibinfo{volume}{9}},
  \bibinfo{pages}{3280} (\bibinfo{year}{2018}).

\bibitem[{\citenamefont{Li et~al.}(2019)\citenamefont{Li, MacDonald, and
  Chen}}]{li2019quantum}
\bibinfo{author}{\bibfnamefont{X.}~\bibnamefont{Li}},
  \bibinfo{author}{\bibfnamefont{A.~H.} \bibnamefont{MacDonald}},
  \bibnamefont{and} \bibinfo{author}{\bibfnamefont{H.}~\bibnamefont{Chen}},
  \bibinfo{journal}{arXiv:1902.10650}  (\bibinfo{year}{2019}).

\bibitem[{\citenamefont{{\v{S}}mejkal et~al.}(2019)\citenamefont{{\v{S}}mejkal,
  Gonz{\'a}lez-Hern{\'a}ndez, Jungwirth, and Sinova}}]{vsmejkal2019crystal}
\bibinfo{author}{\bibfnamefont{L.}~\bibnamefont{{\v{S}}mejkal}},
  \bibinfo{author}{\bibfnamefont{R.}~\bibnamefont{Gonz{\'a}lez-Hern{\'a}ndez}},
  \bibinfo{author}{\bibfnamefont{T.}~\bibnamefont{Jungwirth}},
  \bibnamefont{and} \bibinfo{author}{\bibfnamefont{J.}~\bibnamefont{Sinova}},
  \bibinfo{journal}{arXiv:1901.00445}  (\bibinfo{year}{2019}).

\bibitem[{\citenamefont{Suzuki et~al.}(2018)\citenamefont{Suzuki, Ikeda, and
  Oppeneer}}]{suzuki2018first}
\bibinfo{author}{\bibfnamefont{M.-T.} \bibnamefont{Suzuki}},
  \bibinfo{author}{\bibfnamefont{H.}~\bibnamefont{Ikeda}}, \bibnamefont{and}
  \bibinfo{author}{\bibfnamefont{P.~M.} \bibnamefont{Oppeneer}},
  \bibinfo{journal}{J. Phys. Soc. Jpn.} \textbf{\bibinfo{volume}{87}},
  \bibinfo{pages}{041008} (\bibinfo{year}{2018}).

\bibitem[{\citenamefont{Hayami et~al.}(2018)\citenamefont{Hayami, Yatsushiro,
  Yanagi, and Kusunose}}]{Hayami_PhysRevB.98.165110}
\bibinfo{author}{\bibfnamefont{S.}~\bibnamefont{Hayami}},
  \bibinfo{author}{\bibfnamefont{M.}~\bibnamefont{Yatsushiro}},
  \bibinfo{author}{\bibfnamefont{Y.}~\bibnamefont{Yanagi}}, \bibnamefont{and}
  \bibinfo{author}{\bibfnamefont{H.}~\bibnamefont{Kusunose}},
  \bibinfo{journal}{Phys. Rev. B} \textbf{\bibinfo{volume}{98}},
  \bibinfo{pages}{165110} (\bibinfo{year}{2018}).

\bibitem[{\citenamefont{Watanabe and
  Yanase}(2018)}]{Watanabe_PhysRevB.98.245129}
\bibinfo{author}{\bibfnamefont{H.}~\bibnamefont{Watanabe}} \bibnamefont{and}
  \bibinfo{author}{\bibfnamefont{Y.}~\bibnamefont{Yanase}},
  \bibinfo{journal}{Phys. Rev. B} \textbf{\bibinfo{volume}{98}},
  \bibinfo{pages}{245129} (\bibinfo{year}{2018}).

\bibitem[{\citenamefont{Zhang et~al.}(2018)\citenamefont{Zhang,
  {\v{Z}}elezn{\`y}, Sun, van~den Brink, and Yan}}]{zhang2018spin}
\bibinfo{author}{\bibfnamefont{Y.}~\bibnamefont{Zhang}},
  \bibinfo{author}{\bibfnamefont{J.}~\bibnamefont{{\v{Z}}elezn{\`y}}},
  \bibinfo{author}{\bibfnamefont{Y.}~\bibnamefont{Sun}},
  \bibinfo{author}{\bibfnamefont{J.}~\bibnamefont{van~den Brink}},
  \bibnamefont{and} \bibinfo{author}{\bibfnamefont{B.}~\bibnamefont{Yan}},
  \bibinfo{journal}{New J. Phys.} \textbf{\bibinfo{volume}{20}},
  \bibinfo{pages}{073028} (\bibinfo{year}{2018}).

\bibitem[{\citenamefont{Hayami et~al.}(2019{\natexlab{a}})\citenamefont{Hayami,
  Yanagi, and Kusunose}}]{hayami2019momentum}
\bibinfo{author}{\bibfnamefont{S.}~\bibnamefont{Hayami}},
  \bibinfo{author}{\bibfnamefont{Y.}~\bibnamefont{Yanagi}}, \bibnamefont{and}
  \bibinfo{author}{\bibfnamefont{H.}~\bibnamefont{Kusunose}},
  \bibinfo{journal}{J. Phys. Soc. Jpn.} \textbf{\bibinfo{volume}{88}},
  \bibinfo{pages}{123702} (\bibinfo{year}{2019}{\natexlab{a}}).

\bibitem[{\citenamefont{Naka et~al.}(2019)\citenamefont{Naka, Hayami, Kusunose,
  Yanagi, Motome, and Seo}}]{naka2019spin}
\bibinfo{author}{\bibfnamefont{M.}~\bibnamefont{Naka}},
  \bibinfo{author}{\bibfnamefont{S.}~\bibnamefont{Hayami}},
  \bibinfo{author}{\bibfnamefont{H.}~\bibnamefont{Kusunose}},
  \bibinfo{author}{\bibfnamefont{Y.}~\bibnamefont{Yanagi}},
  \bibinfo{author}{\bibfnamefont{Y.}~\bibnamefont{Motome}}, \bibnamefont{and}
  \bibinfo{author}{\bibfnamefont{H.}~\bibnamefont{Seo}},
  \bibinfo{journal}{Nature communications} \textbf{\bibinfo{volume}{10}}
  (\bibinfo{year}{2019}).

\bibitem[{\citenamefont{Berlijn et~al.}(2017)\citenamefont{Berlijn, Snijders,
  Delaire, Zhou, Maier, Cao, Chi, Matsuda, Wang, Koehler
  et~al.}}]{Berlijn_PhysRevLett.118.077201}
\bibinfo{author}{\bibfnamefont{T.}~\bibnamefont{Berlijn}},
  \bibinfo{author}{\bibfnamefont{P.~C.} \bibnamefont{Snijders}},
  \bibinfo{author}{\bibfnamefont{O.}~\bibnamefont{Delaire}},
  \bibinfo{author}{\bibfnamefont{H.-D.} \bibnamefont{Zhou}},
  \bibinfo{author}{\bibfnamefont{T.~A.} \bibnamefont{Maier}},
  \bibinfo{author}{\bibfnamefont{H.-B.} \bibnamefont{Cao}},
  \bibinfo{author}{\bibfnamefont{S.-X.} \bibnamefont{Chi}},
  \bibinfo{author}{\bibfnamefont{M.}~\bibnamefont{Matsuda}},
  \bibinfo{author}{\bibfnamefont{Y.}~\bibnamefont{Wang}},
  \bibinfo{author}{\bibfnamefont{M.~R.} \bibnamefont{Koehler}},
  \bibnamefont{et~al.}, \bibinfo{journal}{Phys. Rev. Lett.}
  \textbf{\bibinfo{volume}{118}}, \bibinfo{pages}{077201}
  (\bibinfo{year}{2017}).

\bibitem[{\citenamefont{Ahn et~al.}(2019)\citenamefont{Ahn, Hariki, Lee, and
  Kune\ifmmode~\check{s}\else \v{s}\fi{}}}]{Ahn_PhysRevB.99.184432}
\bibinfo{author}{\bibfnamefont{K.-H.} \bibnamefont{Ahn}},
  \bibinfo{author}{\bibfnamefont{A.}~\bibnamefont{Hariki}},
  \bibinfo{author}{\bibfnamefont{K.-W.} \bibnamefont{Lee}}, \bibnamefont{and}
  \bibinfo{author}{\bibfnamefont{J.}~\bibnamefont{Kune\ifmmode~\check{s}\else
  \v{s}\fi{}}}, \bibinfo{journal}{Phys. Rev. B} \textbf{\bibinfo{volume}{99}},
  \bibinfo{pages}{184432} (\bibinfo{year}{2019}).

\bibitem[{\citenamefont{Hayami and Kusunose}(2018)}]{hayami2018microscopic}
\bibinfo{author}{\bibfnamefont{S.}~\bibnamefont{Hayami}} \bibnamefont{and}
  \bibinfo{author}{\bibfnamefont{H.}~\bibnamefont{Kusunose}},
  \bibinfo{journal}{J. Phys. Soc. Jpn.} \textbf{\bibinfo{volume}{87}},
  \bibinfo{pages}{033709} (\bibinfo{year}{2018}).

\bibitem[{\citenamefont{Yoshioka et~al.}(2009)\citenamefont{Yoshioka, Koga, and
  Kawakami}}]{Yoshioka_PhysRevLett.103.036401}
\bibinfo{author}{\bibfnamefont{T.}~\bibnamefont{Yoshioka}},
  \bibinfo{author}{\bibfnamefont{A.}~\bibnamefont{Koga}}, \bibnamefont{and}
  \bibinfo{author}{\bibfnamefont{N.}~\bibnamefont{Kawakami}},
  \bibinfo{journal}{Phys. Rev. Lett.} \textbf{\bibinfo{volume}{103}},
  \bibinfo{pages}{036401} (\bibinfo{year}{2009}).

\bibitem[{\citenamefont{Xiao et~al.}(2012)\citenamefont{Xiao, Liu, Feng, Xu,
  and Yao}}]{xiao2012coupled}
\bibinfo{author}{\bibfnamefont{D.}~\bibnamefont{Xiao}},
  \bibinfo{author}{\bibfnamefont{G.-B.} \bibnamefont{Liu}},
  \bibinfo{author}{\bibfnamefont{W.}~\bibnamefont{Feng}},
  \bibinfo{author}{\bibfnamefont{X.}~\bibnamefont{Xu}}, \bibnamefont{and}
  \bibinfo{author}{\bibfnamefont{W.}~\bibnamefont{Yao}},
  \bibinfo{journal}{Phys. Rev. Lett.} \textbf{\bibinfo{volume}{108}},
  \bibinfo{pages}{196802} (\bibinfo{year}{2012}).

\bibitem[{\citenamefont{Mak et~al.}(2012)\citenamefont{Mak, He, Shan, and
  Heinz}}]{mak2012control}
\bibinfo{author}{\bibfnamefont{K.~F.} \bibnamefont{Mak}},
  \bibinfo{author}{\bibfnamefont{K.}~\bibnamefont{He}},
  \bibinfo{author}{\bibfnamefont{J.}~\bibnamefont{Shan}}, \bibnamefont{and}
  \bibinfo{author}{\bibfnamefont{T.~F.} \bibnamefont{Heinz}},
  \bibinfo{journal}{Nat. Nanotech.} \textbf{\bibinfo{volume}{7}},
  \bibinfo{pages}{494} (\bibinfo{year}{2012}).

\bibitem[{\citenamefont{Zeng et~al.}(2012)\citenamefont{Zeng, Dai, Yao, Xiao,
  and Cui}}]{zeng2012valley}
\bibinfo{author}{\bibfnamefont{H.}~\bibnamefont{Zeng}},
  \bibinfo{author}{\bibfnamefont{J.}~\bibnamefont{Dai}},
  \bibinfo{author}{\bibfnamefont{W.}~\bibnamefont{Yao}},
  \bibinfo{author}{\bibfnamefont{D.}~\bibnamefont{Xiao}}, \bibnamefont{and}
  \bibinfo{author}{\bibfnamefont{X.}~\bibnamefont{Cui}}, \bibinfo{journal}{Nat.
  Nanotech.} \textbf{\bibinfo{volume}{7}}, \bibinfo{pages}{490}
  (\bibinfo{year}{2012}).

\bibitem[{\citenamefont{Suzuki et~al.}(2014)\citenamefont{Suzuki, Sakano,
  Zhang, Akashi, Morikawa, Harasawa, Yaji, Kuroda, Miyamoto, Okuda
  et~al.}}]{suzuki2014valley}
\bibinfo{author}{\bibfnamefont{R.}~\bibnamefont{Suzuki}},
  \bibinfo{author}{\bibfnamefont{M.}~\bibnamefont{Sakano}},
  \bibinfo{author}{\bibfnamefont{Y.}~\bibnamefont{Zhang}},
  \bibinfo{author}{\bibfnamefont{R.}~\bibnamefont{Akashi}},
  \bibinfo{author}{\bibfnamefont{D.}~\bibnamefont{Morikawa}},
  \bibinfo{author}{\bibfnamefont{A.}~\bibnamefont{Harasawa}},
  \bibinfo{author}{\bibfnamefont{K.}~\bibnamefont{Yaji}},
  \bibinfo{author}{\bibfnamefont{K.}~\bibnamefont{Kuroda}},
  \bibinfo{author}{\bibfnamefont{K.}~\bibnamefont{Miyamoto}},
  \bibinfo{author}{\bibfnamefont{T.}~\bibnamefont{Okuda}},
  \bibnamefont{et~al.}, \bibinfo{journal}{Nat. Nanotech.}
  \textbf{\bibinfo{volume}{9}}, \bibinfo{pages}{611} (\bibinfo{year}{2014}).

\bibitem[{\citenamefont{Suzuki et~al.}(2017)\citenamefont{Suzuki, Koretsune,
  Ochi, and Arita}}]{Suzuki_PhysRevB.95.094406}
\bibinfo{author}{\bibfnamefont{M.-T.} \bibnamefont{Suzuki}},
  \bibinfo{author}{\bibfnamefont{T.}~\bibnamefont{Koretsune}},
  \bibinfo{author}{\bibfnamefont{M.}~\bibnamefont{Ochi}}, \bibnamefont{and}
  \bibinfo{author}{\bibfnamefont{R.}~\bibnamefont{Arita}},
  \bibinfo{journal}{Phys. Rev. B} \textbf{\bibinfo{volume}{95}},
  \bibinfo{pages}{094406} (\bibinfo{year}{2017}).

\bibitem[{\citenamefont{Suzuki et~al.}(2019)\citenamefont{Suzuki, Nomoto,
  Arita, Yanagi, Hayami, and Kusunose}}]{Suzuki_PhysRevB.99.174407}
\bibinfo{author}{\bibfnamefont{M.-T.} \bibnamefont{Suzuki}},
  \bibinfo{author}{\bibfnamefont{T.}~\bibnamefont{Nomoto}},
  \bibinfo{author}{\bibfnamefont{R.}~\bibnamefont{Arita}},
  \bibinfo{author}{\bibfnamefont{Y.}~\bibnamefont{Yanagi}},
  \bibinfo{author}{\bibfnamefont{S.}~\bibnamefont{Hayami}}, \bibnamefont{and}
  \bibinfo{author}{\bibfnamefont{H.}~\bibnamefont{Kusunose}},
  \bibinfo{journal}{Phys. Rev. B} \textbf{\bibinfo{volume}{99}},
  \bibinfo{pages}{174407} (\bibinfo{year}{2019}).

\bibitem[{\citenamefont{Ederer and Spaldin}(2007)}]{EdererPhysRevB.76.214404}
\bibinfo{author}{\bibfnamefont{C.}~\bibnamefont{Ederer}} \bibnamefont{and}
  \bibinfo{author}{\bibfnamefont{N.~A.} \bibnamefont{Spaldin}},
  \bibinfo{journal}{Phys. Rev. B} \textbf{\bibinfo{volume}{76}},
  \bibinfo{pages}{214404} (\bibinfo{year}{2007}).

\bibitem[{\citenamefont{Spaldin et~al.}(2008)\citenamefont{Spaldin, Fiebig, and
  Mostovoy}}]{Spaldin_0953-8984-20-43-434203}
\bibinfo{author}{\bibfnamefont{N.~A.} \bibnamefont{Spaldin}},
  \bibinfo{author}{\bibfnamefont{M.}~\bibnamefont{Fiebig}}, \bibnamefont{and}
  \bibinfo{author}{\bibfnamefont{M.}~\bibnamefont{Mostovoy}},
  \bibinfo{journal}{J. Phys.: Condens. Matter} \textbf{\bibinfo{volume}{20}},
  \bibinfo{pages}{434203} (\bibinfo{year}{2008}).

\bibitem[{\citenamefont{Hayami et~al.}(2014)\citenamefont{Hayami, Kusunose, and
  Motome}}]{Hayami_PhysRevB.90.024432}
\bibinfo{author}{\bibfnamefont{S.}~\bibnamefont{Hayami}},
  \bibinfo{author}{\bibfnamefont{H.}~\bibnamefont{Kusunose}}, \bibnamefont{and}
  \bibinfo{author}{\bibfnamefont{Y.}~\bibnamefont{Motome}},
  \bibinfo{journal}{Phys. Rev. B} \textbf{\bibinfo{volume}{90}},
  \bibinfo{pages}{024432} (\bibinfo{year}{2014}).

\bibitem[{\citenamefont{Gao et~al.}(2018)\citenamefont{Gao, Vanderbilt, and
  Xiao}}]{Gao_PhysRevB.97.134423}
\bibinfo{author}{\bibfnamefont{Y.}~\bibnamefont{Gao}},
  \bibinfo{author}{\bibfnamefont{D.}~\bibnamefont{Vanderbilt}},
  \bibnamefont{and} \bibinfo{author}{\bibfnamefont{D.}~\bibnamefont{Xiao}},
  \bibinfo{journal}{Phys. Rev. B} \textbf{\bibinfo{volume}{97}},
  \bibinfo{pages}{134423} (\bibinfo{year}{2018}).

\bibitem[{\citenamefont{Hayami et~al.}(2019{\natexlab{b}})\citenamefont{Hayami,
  Yanagi, Kusunose, and Motome}}]{Hayami_PhysRevLett.122.147602}
\bibinfo{author}{\bibfnamefont{S.}~\bibnamefont{Hayami}},
  \bibinfo{author}{\bibfnamefont{Y.}~\bibnamefont{Yanagi}},
  \bibinfo{author}{\bibfnamefont{H.}~\bibnamefont{Kusunose}}, \bibnamefont{and}
  \bibinfo{author}{\bibfnamefont{Y.}~\bibnamefont{Motome}},
  \bibinfo{journal}{Phys. Rev. Lett.} \textbf{\bibinfo{volume}{122}},
  \bibinfo{pages}{147602} (\bibinfo{year}{2019}{\natexlab{b}}).

\bibitem[{com({\natexlab{a}})}]{comment_symmetric_SS}
\bibinfo{note}{The active E bond multipoles can lead to the symmetric spin
  splitting under magnetic orders~\cite{hayami2019momentum}.}

\bibitem[{com({\natexlab{b}})}]{comment_mpform}
\bibinfo{note}{The band deformations are also obtained by higher-order
  multiplication of E and MT multipoles belonging to the same irreducible
  representation.}

\bibitem[{\citenamefont{Cheon et~al.}(2018)\citenamefont{Cheon, Lee, and
  Cheong}}]{Cheon_PhysRevB.98.184405}
\bibinfo{author}{\bibfnamefont{S.}~\bibnamefont{Cheon}},
  \bibinfo{author}{\bibfnamefont{H.-W.} \bibnamefont{Lee}}, \bibnamefont{and}
  \bibinfo{author}{\bibfnamefont{S.-W.} \bibnamefont{Cheong}},
  \bibinfo{journal}{Phys. Rev. B} \textbf{\bibinfo{volume}{98}},
  \bibinfo{pages}{184405} (\bibinfo{year}{2018}).

\bibitem[{\citenamefont{Maksimov et~al.}(2019)\citenamefont{Maksimov, Zhu,
  White, and Chernyshev}}]{Maksimov_PhysRevX.9.021017}
\bibinfo{author}{\bibfnamefont{P.~A.} \bibnamefont{Maksimov}},
  \bibinfo{author}{\bibfnamefont{Z.}~\bibnamefont{Zhu}},
  \bibinfo{author}{\bibfnamefont{S.~R.} \bibnamefont{White}}, \bibnamefont{and}
  \bibinfo{author}{\bibfnamefont{A.~L.} \bibnamefont{Chernyshev}},
  \bibinfo{journal}{Phys. Rev. X} \textbf{\bibinfo{volume}{9}},
  \bibinfo{pages}{021017} (\bibinfo{year}{2019}).

\bibitem[{\citenamefont{Hayashida et~al.}(2018)\citenamefont{Hayashida,
  Zaharko, Kurita, Tanaka, Hagihala, Soda, Itoh, Uwatoko, and
  Masuda}}]{Hayashida_PhysRevB.97.140405}
\bibinfo{author}{\bibfnamefont{S.}~\bibnamefont{Hayashida}},
  \bibinfo{author}{\bibfnamefont{O.}~\bibnamefont{Zaharko}},
  \bibinfo{author}{\bibfnamefont{N.}~\bibnamefont{Kurita}},
  \bibinfo{author}{\bibfnamefont{H.}~\bibnamefont{Tanaka}},
  \bibinfo{author}{\bibfnamefont{M.}~\bibnamefont{Hagihala}},
  \bibinfo{author}{\bibfnamefont{M.}~\bibnamefont{Soda}},
  \bibinfo{author}{\bibfnamefont{S.}~\bibnamefont{Itoh}},
  \bibinfo{author}{\bibfnamefont{Y.}~\bibnamefont{Uwatoko}}, \bibnamefont{and}
  \bibinfo{author}{\bibfnamefont{T.}~\bibnamefont{Masuda}},
  \bibinfo{journal}{Phys. Rev. B} \textbf{\bibinfo{volume}{97}},
  \bibinfo{pages}{140405} (\bibinfo{year}{2018}).

\bibitem[{\citenamefont{Mekata et~al.}(1995)\citenamefont{Mekata, Sugino,
  Oohara, Oohara, and Yoshizawa}}]{mekata1995magnetic}
\bibinfo{author}{\bibfnamefont{M.}~\bibnamefont{Mekata}},
  \bibinfo{author}{\bibfnamefont{T.}~\bibnamefont{Sugino}},
  \bibinfo{author}{\bibfnamefont{A.}~\bibnamefont{Oohara}},
  \bibinfo{author}{\bibfnamefont{Y.}~\bibnamefont{Oohara}}, \bibnamefont{and}
  \bibinfo{author}{\bibfnamefont{H.}~\bibnamefont{Yoshizawa}},
  \bibinfo{journal}{Physica B: Condens. Matter} \textbf{\bibinfo{volume}{213}},
  \bibinfo{pages}{221} (\bibinfo{year}{1995}).

\bibitem[{\citenamefont{Ghannadzadeh et~al.}(2017)\citenamefont{Ghannadzadeh,
  Licciardello, Arsenijevi{\'c}, Robinson, Takatsu, Katsnelson, and
  Hussey}}]{ghannadzadeh2017simultaneous}
\bibinfo{author}{\bibfnamefont{S.}~\bibnamefont{Ghannadzadeh}},
  \bibinfo{author}{\bibfnamefont{S.}~\bibnamefont{Licciardello}},
  \bibinfo{author}{\bibfnamefont{S.}~\bibnamefont{Arsenijevi{\'c}}},
  \bibinfo{author}{\bibfnamefont{P.}~\bibnamefont{Robinson}},
  \bibinfo{author}{\bibfnamefont{H.}~\bibnamefont{Takatsu}},
  \bibinfo{author}{\bibfnamefont{M.}~\bibnamefont{Katsnelson}},
  \bibnamefont{and} \bibinfo{author}{\bibfnamefont{N.}~\bibnamefont{Hussey}},
  \bibinfo{journal}{Nat. Commun.} \textbf{\bibinfo{volume}{8}},
  \bibinfo{pages}{15001} (\bibinfo{year}{2017}).

\bibitem[{\citenamefont{Foggetti et~al.}(2019)\citenamefont{Foggetti, Cheong,
  and Artyukhin}}]{Foggetti_PhysRevB.100.180408}
\bibinfo{author}{\bibfnamefont{F.}~\bibnamefont{Foggetti}},
  \bibinfo{author}{\bibfnamefont{S.-W.} \bibnamefont{Cheong}},
  \bibnamefont{and}
  \bibinfo{author}{\bibfnamefont{S.}~\bibnamefont{Artyukhin}},
  \bibinfo{journal}{Phys. Rev. B} \textbf{\bibinfo{volume}{100}},
  \bibinfo{pages}{180408} (\bibinfo{year}{2019}).

\bibitem[{\citenamefont{Seki and Okunishi}(2015)}]{Seki_PhysRevB.91.224403}
\bibinfo{author}{\bibfnamefont{K.}~\bibnamefont{Seki}} \bibnamefont{and}
  \bibinfo{author}{\bibfnamefont{K.}~\bibnamefont{Okunishi}},
  \bibinfo{journal}{Phys. Rev. B} \textbf{\bibinfo{volume}{91}},
  \bibinfo{pages}{224403} (\bibinfo{year}{2015}).

\bibitem[{\citenamefont{Hagihala et~al.}(2019)\citenamefont{Hagihala,
  Hayashida, Avdeev, Manaka, Kikuchi, and Masuda}}]{hagihala2019magnetic}
\bibinfo{author}{\bibfnamefont{M.}~\bibnamefont{Hagihala}},
  \bibinfo{author}{\bibfnamefont{S.}~\bibnamefont{Hayashida}},
  \bibinfo{author}{\bibfnamefont{M.}~\bibnamefont{Avdeev}},
  \bibinfo{author}{\bibfnamefont{H.}~\bibnamefont{Manaka}},
  \bibinfo{author}{\bibfnamefont{H.}~\bibnamefont{Kikuchi}}, \bibnamefont{and}
  \bibinfo{author}{\bibfnamefont{T.}~\bibnamefont{Masuda}},
  \bibinfo{journal}{npj Quantum Mater.} \textbf{\bibinfo{volume}{4}},
  \bibinfo{pages}{14} (\bibinfo{year}{2019}).

\bibitem[{\citenamefont{Lee et~al.}(2014)\citenamefont{Lee, Choi, Huang, Ma,
  Dela~Cruz, Matsuda, Tian, Dun, Dong, and Zhou}}]{Lee_PhysRevB.90.224402}
\bibinfo{author}{\bibfnamefont{M.}~\bibnamefont{Lee}},
  \bibinfo{author}{\bibfnamefont{E.~S.} \bibnamefont{Choi}},
  \bibinfo{author}{\bibfnamefont{X.}~\bibnamefont{Huang}},
  \bibinfo{author}{\bibfnamefont{J.}~\bibnamefont{Ma}},
  \bibinfo{author}{\bibfnamefont{C.~R.} \bibnamefont{Dela~Cruz}},
  \bibinfo{author}{\bibfnamefont{M.}~\bibnamefont{Matsuda}},
  \bibinfo{author}{\bibfnamefont{W.}~\bibnamefont{Tian}},
  \bibinfo{author}{\bibfnamefont{Z.~L.} \bibnamefont{Dun}},
  \bibinfo{author}{\bibfnamefont{S.}~\bibnamefont{Dong}}, \bibnamefont{and}
  \bibinfo{author}{\bibfnamefont{H.~D.} \bibnamefont{Zhou}},
  \bibinfo{journal}{Phys. Rev. B} \textbf{\bibinfo{volume}{90}},
  \bibinfo{pages}{224402} (\bibinfo{year}{2014}).

\bibitem[{\citenamefont{Perdew et~al.}(1996)\citenamefont{Perdew, Burke, and
  Ernzerhof}}]{Perdew_PhysRevLett.77.3865}
\bibinfo{author}{\bibfnamefont{J.~P.} \bibnamefont{Perdew}},
  \bibinfo{author}{\bibfnamefont{K.}~\bibnamefont{Burke}}, \bibnamefont{and}
  \bibinfo{author}{\bibfnamefont{M.}~\bibnamefont{Ernzerhof}},
  \bibinfo{journal}{Phys. Rev. Lett.} \textbf{\bibinfo{volume}{77}},
  \bibinfo{pages}{3865} (\bibinfo{year}{1996}).

\bibitem[{\citenamefont{Dudarev et~al.}(1998)\citenamefont{Dudarev, Botton,
  Savrasov, Humphreys, and Sutton}}]{Dudarev_PhysRevB.57.1505}
\bibinfo{author}{\bibfnamefont{S.~L.} \bibnamefont{Dudarev}},
  \bibinfo{author}{\bibfnamefont{G.~A.} \bibnamefont{Botton}},
  \bibinfo{author}{\bibfnamefont{S.~Y.} \bibnamefont{Savrasov}},
  \bibinfo{author}{\bibfnamefont{C.~J.} \bibnamefont{Humphreys}},
  \bibnamefont{and} \bibinfo{author}{\bibfnamefont{A.~P.}
  \bibnamefont{Sutton}}, \bibinfo{journal}{Phys. Rev. B}
  \textbf{\bibinfo{volume}{57}}, \bibinfo{pages}{1505} (\bibinfo{year}{1998}).

\bibitem[{\citenamefont{Kresse and
  Furthm\"uller}(1996)}]{Kresse_PhysRevB.54.11169}
\bibinfo{author}{\bibfnamefont{G.}~\bibnamefont{Kresse}} \bibnamefont{and}
  \bibinfo{author}{\bibfnamefont{J.}~\bibnamefont{Furthm\"uller}},
  \bibinfo{journal}{Phys. Rev. B} \textbf{\bibinfo{volume}{54}},
  \bibinfo{pages}{11169} (\bibinfo{year}{1996}).

\bibitem[{vas()}]{vasp}
\bibinfo{note}{Vienna {\it Ab initio} Simulation Package, http://www.vasp.at.}

\bibitem[{\citenamefont{Bl\"ochl}(1994)}]{Blochl_PhysRevB.50.17953}
\bibinfo{author}{\bibfnamefont{P.~E.} \bibnamefont{Bl\"ochl}},
  \bibinfo{journal}{Phys. Rev. B} \textbf{\bibinfo{volume}{50}},
  \bibinfo{pages}{17953} (\bibinfo{year}{1994}).

\bibitem[{\citenamefont{Kresse and Joubert}(1999)}]{Kresse_PhysRevB.59.1758}
\bibinfo{author}{\bibfnamefont{G.}~\bibnamefont{Kresse}} \bibnamefont{and}
  \bibinfo{author}{\bibfnamefont{D.}~\bibnamefont{Joubert}},
  \bibinfo{journal}{Phys. Rev. B} \textbf{\bibinfo{volume}{59}},
  \bibinfo{pages}{1758} (\bibinfo{year}{1999}).

\end{thebibliography}

\end{document}